\documentclass{ws-ijmpd}

\usepackage[super,compress]{cite}
\usepackage{graphicx}
\usepackage{epstopdf}
\usepackage{amsfonts}
\usepackage{amssymb}
\usepackage{epsfig}
\usepackage{hyperref}
\usepackage{color}

\def\de#1/de#2{\frac{\partial {#1}}{\partial {#2}}}

\newcommand{\ba}{\begin{eqnarray}}
\newcommand{\ea}{\end{eqnarray}}
\newcommand{\be}{\begin{equation}}
\newcommand{\ee}{\end{equation}}

\begin{document}


%
\catchline{}{}{}{}{}
%

\title{On the impact of non-local gravity on compact stars}

\author{Grigoris Panotopoulos}

\address{
Centro de Astrof{\'i}sica e Gravita{\c c}{\~a}o-CENTRA, Instituto Superior T{\'e}cnico-IST, Universidade de Lisboa-UL, Av. Rovisco Pais, 1049-001 Lisboa, Portugal \\
Departamento de Ciencias F{\'i}sicas, Universidad de la Frontera, Casilla 54-D, 4811186 Temuco, Chile.
\\
\href{mailto:grigorios.panotopoulos@ufrontera.cl}{\nolinkurl{grigorios.panotopoulos@ufrontera.cl}} 
}

\author{Javier Rubio}

\address{
Centro de Astrof{\'i}sica e Gravita{\c c}{\~a}o-CENTRA, Instituto Superior T{\'e}cnico-IST, Universidade de Lisboa-UL, Av. Rovisco Pais, 1049-001 Lisboa, Portugal.\\ 
Departamento de F\'isica Te\'orica and Instituto de F\'isica de Part\'iculas y del Cosmos (IPARCOS-UCM), Universidad Complutense de Madrid, 28040 
Madrid, Spain
\\
\href{mailto:javier.rubio@ucm.es}{\nolinkurl{javier.rubio@ucm.es}} 
}

\author{Il{\'i}dio Lopes}

\address{
Centro de Astrof{\'i}sica e Gravita{\c c}{\~a}o-CENTRA, Instituto Superior T{\'e}cnico-IST, Universidade de Lisboa-UL, Av. Rovisco Pais, 1049-001 Lisboa, Portugal.
\\
\href{mailto:ilidio.lopes@tecnico.ulisboa.pt}{\nolinkurl{ilidio.lopes@tecnico.ulisboa.pt}} 
}

\maketitle

\begin{history}
\received{Day Month Year}
\revised{Day Month Year}
\end{history}

\begin{abstract}
We study the impact of non-local modifications of General Relativity on stellar structure. In particular, assuming an analytic distortion function and specific equations of state, we made use of remnant stars to put qualitative constraints on a parameter not directly restricted by solar system tests. Using current data sets available for white dwarfs and strange quark stars candidates we find that the most stringent bounds come from the objects displaying the highest core densities, namely strange quark stars. Specifically, the constraints obtained from this class of stars are three to four orders of magnitude tighter than those obtained using white dwarfs.
\end{abstract}


\ccode{}

\section{Introduction}\label{sec:introduction}

From the very inception of General Relativity (GR) people have looked for alternative theories of gravity. Most of these extensions are manifestly local. Note, however,  that non-localities may appear naturally at high and low energies. The possibility that gravitational interactions become non-local near the Planck scale is suggested among others by string theory \cite{Siegel:2003vt,Calcagni:2014vxa}. On the other hand, infrared non-localities may appear in effective field theories obtained by integrating out light degrees of freedom \cite{Barvinsky:2014lja,Codello:2015mba,Donoghue:2015nba} or in quantum gravity approaches aimed to solve the unboundedness of the Euclidean Einstein-Hilbert action \cite{Wetterich:1997bz,Barvinsky:2011rk}.  In particular, since graviton fluctuations exists generically around flat spacetimes, there is no fundamental reason to assume that the effective action for gravity must be local \cite{Wetterich:1997bz}.

The implications on non-localities at large scales have been extensively studied in the literature, with special emphasis on the early and late acceleration of the Universe  \cite{Deser:2007jk,Nojiri:2007uq,Capozziello:2008gu,Koivisto:2008dh,Deffayet:2009ca,Biswas:2010zk,Nojiri:2010wj,Biswas:2012bp,Deser:2013uya,Woodard:2014iga,ArkaniHamed:2002fu,Dvali:2006su,Dvali:2007kt,Elizalde:2012ja,Maggiore:2013mea,Barreira:2014kra,Conroy:2014eja,Nersisyan:2016hjh,Chen:2019wlu,Nersisyan:2016jta,Belgacem:2018wtb,Amendola:2019fhc,Deser:2019lmm,Tian:2019bla}. In this work, however, we turn our attention to potential small-scale constraints coming from idealized spherically symmetric remnant stars, without invoking any extrapolation of our results to cosmological scales.
This type of approach has been used to constrain the parameter space of different modified gravity theories, ranging from scalar-tensor actions of the Horndeski type \cite{Koyama:2015oma,Jain:2015edg,2018JCAP...05..028S,Jain:2015edg} to $f(R)$ models \cite{Carvalho:2019gzs} or exotic ungravity scenarios \cite{Bertolami:2016ylu}.\footnote{Note, however, that, contrary to the approach followed in this paper, some of these references assume the constraints derived at small scales to be valid also in a cosmological context.} For the sake of concreteness, we will focus here on a specific setting in which the effect of non-localities is encoded on functionals of the dimensionless covariant combination $\phi\equiv \Box^{-1} R$, with $R$ the Ricci scalar and $\Box^{-1}$ the Green function of the d'Alembertian operator $\Box\equiv \nabla^{\mu}\nabla_{\mu}$, with $\nabla_{\mu}$ the Christoffel covariant derivative \cite{Wetterich:1997bz}. For a given spherical collapsed structure of mass $M$ and size $L$, this quantity is of the order of the \textit{compactness}, defined as the ratio of Schwarzschild radius to the radius of the object, $\phi \equiv  R/\Box  \sim  R\times L^2 \sim G M/L$. While this parameter is significantly smaller than one in the solar system, it becomes sizable in more compact objects such as white dwarfs, neutron stars, or strange quark stars, opening the possibility of testing non-local gravity theories with astrophysical observations. 

Obtaining accurate and model-independent astrophysical constraints on modified gravity scenarios is of course a rather subtle issue, since the uncertainties associated with the choice of an equation of state and the modeling of the wide range of densities, temperatures, isospin asymmetries and layers encountered in the stellar interior unavoidably interfere with the assumed theory of gravity. This limitation is particularly relevant for neutron stars since current terrestrial experiments are unable to test the behavior of cold nuclear matter at densities above the saturation density of nuclei, $\rho_{\rm s}= 2.8\times 10^{14} \,\textrm{g}\cdot \textrm{cm}^{-3}$ \cite{Tsang:2012se}, and therefore the extreme conditions taking place in the deep core of these objects. In the lack of a definitive characterization of the equation of state of compact objects, we will restrict ourselves to a qualitative rather than quantitative analysis of the impact on non-localities in compact stars. Nevertheless, we expect the forthcoming gravitational wave, X-ray, and gamma-ray detectors to make compact stars a powerful tool to constrain non-local theories of gravity. In particular, since those stars are scattered across the Milky Way, and more noticeably near the supermassive black hole at the Galactic center, they can validate the theory in different environments within the Galaxy. Moreover, such bounds could be particularly insightful once we have a better knowledge of those stars' inner structure, including their equation of state (EoS). Here, we present a first attempt to constrain a non-local gravity theory using two fiducial classes of compact stars, namely white dwarfs and strange quark stars.

Our work in this article is organized as follows: Section \ref{sec:non-localmodel} introduces non-local theories of gravity by considering an exemplary setting not involving new dimensional parameters beyond those already present in the usual Einstein-Hilbert action. The generalized Tolman-Oppenheimer-Volkoff equations describing the evolution of spherically collapsed systems are presented in Section \ref{TOV}, leaving for Sections \ref{sec:dwarfs} and \ref{sec:strange} their application to specific scenarios such as white dwarfs and strange quark stars.
Finally, our conclusions are presented in Section \ref{sec:conclusions}.

\section{An exemplary non-local model} \label{sec:non-localmodel}

Non-local models of gravity are typically written down as an Einstein-Hilbert term supplemented by a number of integral or infinite-derivative curvature operators \cite{Belgacem:2017cqo}. For the sake of concreteness, we will here consider a simple non-local model
\begin{eqnarray}
\label{actionNL}
S=\int d^4 x \sqrt{-g}\Big\lbrace
\frac{1}{2\kappa^2}R\left[1 + f(\Box^{-1}R)\right]\Big\rbrace +S_{\rm M}
\end{eqnarray}
modifying GR by an arbitrary function $f$ of the dimensionless quantity $\Box^{-1}R$, with $\kappa=M_P^{-1}=(8 \pi G)^{1/2}$ the inverse Planck mass and $S_M$ the matter action. 

In order to study the inhomogeneous response of the \textit{distortion function} $f(\Box^{-1}R)$ to the energy-momentum tensor of compact objects, it is convenient to introduce an auxiliary scalar field $\phi$ and a Lagrange multiplier $\xi$ such that the non-local action \eqref{actionNL} is brought to a local form  \cite{Nojiri:2007uq,Koshelev:2008ie} 
\begin{eqnarray} \label{actionL}
S&=&\int d^4 x \sqrt{-g}\Big\lbrace
\frac{1}{2\kappa^2}\left[R\left(1 + f(\phi)\right)
+ \xi\left(R-\Box\phi\right) \right]
\Big\rbrace +S_{\rm M} \nonumber \\
&=&\int d^4x\sqrt{-g}\Big\lbrace \frac{1}{2\kappa^2} \left[R(1+f(\phi)+\xi )+g^{\mu \nu}\partial_\mu\xi\,\partial_\nu\phi\right)\Big\rbrace+S_{\rm M}\,,
\end{eqnarray}
where in the last step we have integrated out a total derivative.  The associated equations of motion are derived by performing the variation with respect to $\xi$, $\phi$ and $g_{\mu\nu}$, getting respectively 
\begin{eqnarray}  
&&\hspace{-5mm} \Box \phi=R\,, \label{eqphi}\\ 
&&\hspace{-5mm}\Box\,\xi=f_{,\phi}(\phi)R\,,\label{eqxi}\\ 
&&\hspace{-5mm} G_{\mu \nu}(1+f(\phi)+\xi)+(g_{\mu \nu}\Box-\nabla_\mu\nabla_\nu)(f(\phi)+\xi)-\frac{1}{2}g_{\mu \nu}\partial^\rho\xi\partial_\rho\phi
+\partial_{(\mu}\xi\partial_{\nu)}\phi=\kappa^2 T_{\mu \nu}\,,\label{eqg}
\end{eqnarray}
with $T_{\mu\nu}\equiv -2/\sqrt{-g}  \delta S_M/\delta g^{\mu\nu}$ the energy-momentum tensor of the matter action  and the parenthesis around indexes denoting symmetrization. The original non-local form \eqref{actionNL} is recovered by inserting the constraint \eqref{eqphi} into the local action \eqref{actionL}, being therefore the two representations equivalent on-shell. 

The implications of the dual  actions \eqref{actionNL} and \eqref{actionL} in the low-curvature long-wavelength regime have been extensively studied in the cosmological literature, aiming to explain the late-time acceleration of the Universe without a cosmological constant \cite{Deser:2007jk,Nojiri:2007uq,Koivisto:2008dh,Deffayet:2009ca,Nojiri:2010wj,Elizalde:2012ja,Deser:2013uya,Woodard:2014iga,Barreira:2014kra,Conroy:2014eja,Park:2016jym,Belgacem:2018wtb,Chen:2019wlu,Amendola:2019fhc,Deser:2019lmm,Tian:2019bla}. The naive extrapolation of these studies to the astrophysical scenario under consideration is subject, however, to some formal and technical limitations:

\begin{enumerate}
\item  In the lack of a fundamental background-independent determination of the nonlocal effective action, the choice of a specific distortion function should be understood as a mere parametrization of the effect of non-localities \textit{at the considered scales}. This is, implicitly, the approach followed in Refs.~\cite{Deffayet:2009ca,Park:2016jym}, where a specific form of $f$ was engineered to reproduce the background expansion of  $\Lambda$CDM and $w$CDM cosmologies. The extrapolation of this fitting formula to other gravitational backgrounds and scales is clearly not supported by effective field theory arguments. 
\item Even if a given form of the distortion function is assumed to be valid at all scales, the connection between cosmological and astrophysical backgrounds is far from trivial. In this context, one of the most commonly emphasized drawbacks of the action \eqref{actionNL} is the apparent absence of a screening mechanism able to reconcile the temporal variation of the effective Newton constant in scenarios supporting the accelerated expansion of the Universe with the strict bounds coming from Lunar Laser Ranging experiments \cite{Belgacem:2018wtb},  $\dot G_{\rm  eff}/G =(7.1\pm 7.6)\times 10^{-14} \, {\rm yr}^{-1}$ \cite{Hofmann:2018myc}. Albeit this possibility has led to new nonlocal proposals which are interesting on their own \cite{Deser:2019lmm}, it is important to emphasize that the results of Ref.~\cite{Belgacem:2018wtb} are based on simple extrapolations of averaged cosmological observables to solar system scales, namely, on a McVittie-type metric describing the gravity field of a point mass in an expanding Universe \cite{McVittie:1933zz}. Therefore, they do not a priori account for potential nonlinear time- and scale-dependent effects, which are otherwise known to play a key role in the development of screening mechanisms;\footnote{The paradigmatic example is the Vainshtein mechanism \cite{Vainshtein:1972sx}, where is not enough to establish the existence of two asymptotic regimes (the Vainshtein one, where General Relativity is recovered, and the one far from the source where linear Fierz-Pauli massive gravity is recovered) but also to ensure that there exists an everywhere non-singular solution that matches them \cite{Babichev:2013usa}.} cf.~Ref.~\cite{Amendola:2019fhc} for a similar point of view. 
\end{enumerate}
Having the above two arguments in mind, we will assume the distortion function in \eqref{actionL} to be just a convenient parametrization of nonlocalities at the \textit{small scales} under consideration. For the sake of concreteness, we will consider a simple analytic expansion,
\begin{equation}\label{fexp}
f(\phi)=\sum_{i=1}^\infty f_i \,\phi^i
\,,
\end{equation}
with constants coefficients $f_{i>1}$.~\footnote{The omitted constant term $f_0$ may be always absorbed into a redefinition of the effective Planck mass in Eq.~\eqref{actionNL}.} Note, again, that we make no claims about the validity of this parameterization at cosmological distances neither we try to explain the current accelerated expansion of the Universe through it. Indeed, as shown in Ref.~\cite{Koivisto:2008dh}, the terms $\propto f_1$ or  $\propto f_2$ do not lead to viable explanations for the phase of late-time cosmic acceleration, meaning that, if naively extrapolated to cosmological distances, our model should be eventually complemented with a cosmological constant.

It is finally important to note that, in spite of the appearances, we are not dealing with a standard scalar-tensor theory involving new degrees of freedom beyond the metric. In particular, the use of the inverse operator $\Box^{-1}$ in Eq.~\eqref{actionNL} implies a choice of boundary conditions, understood as a property of the theory itself rather than as free parameters to be varied for each solution \cite{Belgacem:2017cqo}. The condition for $\phi$ can be obtained by expanding the Ricci scalar in an eigenvector basis of the D'Alambertian operator \cite{Wetterich:1997bz}, 
\begin{equation}
R(x)=\sum_n a_n R_n(x)\,,\hspace{10mm} \Box \, R_n = \lambda_n R_n(x)\,, \hspace{10mm} \Box^{-1} R(x)=\sum_n a_n \lambda_n^{-1} R_n(x)\,,
\end{equation}
and requiring regularity for non-vanishing eigenvalues $\lambda_n$ ($\Box \,R=0 \Rightarrow  R=0$). For the Ricci-flat geometries we will be interested in in this paper, this implies $\phi=\Box^{-1}R=0$ there \cite{Wetterich:1997bz,Koivisto:2008dh}. We will come back to this condition in the next Section, when describing the exterior of isolated stellar objects [cf.~Eq.~\eqref{fieldboundaries}].  

\section{Tolman-Oppenheimer-Volkoff equations} \label{TOV}

The field equations \eqref{eqphi}-\eqref{eqg} can be reduced to ordinary coupled differential equations by considering a specific spacetime geometry.  Neglecting  for simplicity the stellar rotation, we assume here a static and spherically symmetric line element
\begin{eqnarray}\label{metric}
ds^2=-e^{2\nu(r)}dt^2+e^{2\lambda(r)}dr^2+r^2 (d\theta^2 +\sin^2\theta d\varphi^2) \,,
\end{eqnarray}
with $t$, $r$, $\theta$ and $\varphi$ a set of Schwarzschild-like coordinates and $\nu(r)$ and $\lambda(r)$ two independent functions of the radial coordinate only. Given a perfect energy-momentum tensor $T_{\mu \nu}={\rm diag}(-\rho,p,p,p)$ with $\rho$ the energy density and $p$ the isotropic pressure, the $tt$ and $rr$ components of Einstein equations \eqref{eqg} take the form \cite{Momeni:2015vwa}
\begin{eqnarray}
&&\hspace{-5mm} (1+f(\nu)+\xi)\left(\frac{1-e^{2\lambda}}{r^2}-\frac{2\lambda'}{r}\right)+\frac{1}{2}\xi'\nu'-\left(\frac{2}{r}-\lambda'\right)(f'+\xi')-\left(\phi''f_{\phi}
+\phi'^2f_{\phi\phi}+\xi''\right)=\kappa^2 e^{2\lambda}\rho\,, \nonumber \\ 
&& \hspace{-5mm}  (1+f(\nu)+\xi)\left(\frac{1-e^{2\lambda}}{r^2}+\frac{2\nu'}{r}\right)-\frac12 \xi'\nu'-\left(\frac{2}{r}+\nu'\right)(f'+\xi')=-\kappa^2e^{2\lambda}p
 \label{rr}\,,
\end{eqnarray}
with the primes denoting derivatives with respect to the radial coordinate $r$. Note that the isometries of the metric are inherited by the scalar fields $\phi$ and $\xi$, as well as by the functions constructed out of them. These formulas are supplemented by the field equations \eqref{eqphi} and \eqref{eqxi}, 
\begin{equation}
 \phi''+\Big(\frac{2}{r}+\nu'-\lambda'\Big)\phi'-{\cal G}=0\,, \hspace{10mm}
\xi''+\Big(\frac{2}{r}+\nu'-\lambda'\Big)\xi'-2f_{\phi} \, {\cal G}=0\,, \label{eq:xiM}
\end{equation}
and the trace of Einstein's equations 
\begin{equation}
2(1+f(\phi)+\xi) \, {\cal G}+\phi'\xi'-3\left[(\phi''f_{\phi}
+\phi'^2f_{\phi\phi}+\xi'')+\left(\frac{2}{r}+\nu'-\lambda'\right)(f'+\xi')\right]=\kappa^2 e^{2\lambda}(\rho -p)\,,
\end{equation}
where, for the sake of compactness, we have defined a combination
\begin{equation}
 {\cal G}\equiv \nu''+\nu'^2-\nu'\lambda'+\frac{2}{r}(\nu'-\lambda')+\frac{1-e^{2\lambda}}{r^2}\,.
\end{equation}
Given a distortion function $f(\phi)$, the numerical solutions of this highly non-linear Tolman-Oppenheimer-Volkoff (TOV) problem \cite{PhysRev.55.364,PhysRev.55.374} can be found by assuming either a specific solution for the metric tensor or an equation of state (EoS) $p(\rho)$ characterizing the relationship between the internal pressure and the energy density. The first approach was followed in Ref. \cite{Momeni:2015vwa}, where the authors constructed an empirical equation of state by imposing the metric potentials $\lambda(r)$ and $\nu(r)$ to be quadratic in the radial coordinate $r$. Here, on the contrary, we will assume a physically-motivated EoS right from the beginning and determine from it the metric potentials dictating the observable properties of the star. 

For the distortion function, we will assume the analytic expansion in Eq.~\eqref{fexp} and benefit from the existing solar system constraint on the first term, $\vert f_1\vert \lesssim 5.7 \times 10^{-6}$ \cite{Koivisto:2008dh} to focus our attention in the next coefficient $f_2$ while setting all coefficients $f_{i>2}$ to zero for the sake of simplicity.~\footnote{Note that the solar system constraint on $f_1$ was derived by assuming all higher corrections to be negligible as compared to the linear term \cite{Koivisto:2008dh}. For a compactness parameter $\phi<1$, this requirement translates into specific conditions for the Taylor coefficients in Eq.~\eqref{fexp}. A rough upper limit on the coefficient $f_2$ can be obtained by requiring the consistency of the perturbative treatment. In particular, if the condition  $\vert f_2 \,\phi^2\vert  \leq \vert  f_1\,  \phi\vert$ is not satisfied, higher-order corrections will overpass the first-order contribution, giving rise to inconsistencies with the above solar system tests. Taking into account the value of the compactness parameter in the solar system, $\phi \sim G M_\odot/L_\odot \simeq 2.12 \times 10^{-6}$, we get 
$f_2\lesssim  2.7$, as a first estimate.} For the equation-of-state, we will consider two illustrative cases applicable to compact objects: i) a Chandrasekhar EoS describing white dwarfs and ii) a color-flavor-locked EoS describing strange quark stars. For each of these scenarios, we will numerically integrate the structure equations from the center to the surface of the star, imposing boundary conditions for the fluid, the metric potentials, and the scalar fields at its center, namely
\begin{equation}
m(0)  =  0\,,  \hspace{15mm} \rho(0) = \rho_c\,, \hspace{15mm}
\nu(0)  =  \nu_c\,, \hspace{15mm} \nu'(0) = 0 \,,
\end{equation}
and
\begin{equation} 
\vert \xi(0)\vert    =   5.5 \times 10^{-6} \,, \hspace{15mm} \xi'(0)=0\,,
 \hspace{15mm} \phi(0)  =  \phi_c\,, \hspace{15mm} \phi'(0)=0\,, 
\end{equation}
with the subindex $c$ denoting central values. This provides, among other observables, the mass function of the star,
\begin{equation}
e^{-2 \lambda(r)} \equiv 1-\frac{2\,G\,m(r)}{r}\,,
\end{equation}
as a function of the radial coordinate $r$. Finally, we match the interior solution to a exterior vacuum solution described by the Schwarzschild geometry
\begin{equation}\label{Schmetric}
ds^2=-\left(1-\frac{2GM}{r}\right)dt^2+\left(1-\frac{2GM}{r}\right)^{-1} dr^2+r^2 (d\theta^2 +\sin^2\theta d\varphi^2) \,,
\end{equation}
with $M$ the stellar mass. The matching conditions require (cf. the discussion at the end of Section \ref{sec:non-localmodel})
\begin{equation}
p(R)  =  0\,, \hspace{7mm} \phi(R)  =  0\,,  \hspace{7mm} m(R)  =  M\,,   \hspace{7mm} e^{2 \nu(R)}  =  1-\frac{2GM}{R}\,, \label{fieldboundaries}
\end{equation}
with $R$ the radius of the star. The obtained profiles are then checked to satisfy several physicality requirements, such as the causality condition for the speed of sound,
\begin{equation}\label{cs}
0 < c_s^2 \equiv \frac{dp}{d \rho} < 1\,,
\end{equation}
the stability bound for the adiabatic index \cite{Moustakidis:2016ndw},
\begin{equation}\label{gamma}
\Gamma \equiv c_s^2 \left( 1 + \frac{\rho}{p} \right) >\frac43 \,,
\end{equation} 
and the strong energy condition, 
\begin{equation}\label{SEC}
\rho + p \geq  0\,, \hspace{15mm}\rho + 3\,p \geq 0\,.
\end{equation} 
Note that, if the latest requirement is fulfilled, the weak energy condition ($\rho\geq 0$, $\rho +p \geq 0$), the null energy condition ($\rho +p\geq 0$) and the dominant energy condition ($\rho\geq \vert p\vert$) are automatically satisfied.

Finally, we stress that for a given central pressure and energy density, the central values of $\phi_c$ and $\nu_c$ cannot be picked up at random. Instead, they must be computed by applying a shooting method, such that all the matching conditions are simultaneously fulfilled, and the whole numerical scheme is self-consistent.

\section{White dwarfs} \label{sec:dwarfs}

In a nutshell, the evolution of a white dwarf can be described as the end stage of the evolution of a progenitor main-sequence star with a mass smaller than $10\, M_\odot$. In the advanced post-main sequence phase, the star ejects its outer layers of light elements in the form of a planetary nebula, leaving behind a white dwarf: an Earth-size solar-mass remnant with a Helium or Carbon/Oxygen core and a hydrogen or helium atmosphere \cite{2010AARv..18..471A}. Given the lack of enough light elements to ignite further fusion reactions, the gravitational collapse of this compact star is only prevented by the degeneracy pressure of its constituent electrons. Once formed, the white dwarf cooling is inevitable, starting with an intense neutrino emission of a very hot and  young white dwarf, and ending up as a cold and old white dwarf with a crystallized core \cite{2003ARAA..41..465H}.

While it is nowadays possible to develop quite robust models of white dwarfs accounting for stellar temperature and core composition (see e.g. \cite{Barstow} and \cite{2018JCAP...05..028S} and references therein), we will opt here for a simple zero-temperature model that has the advantage of being analytical and widely used in the literature (see Refs. \cite{1952MNRAS.112..583M} and \cite{Panotopoulos:2020wug}). In particular, in investigating the impact of nonlocalities on the mass-to-radius relationship, we shall assume an ideal Fermi gas with energy density and pressure
\begin{eqnarray}
\rho_{\rm WD}=\frac{2}{(2\pi)^3} \int_{k=0}^{k_F} 4\pi k^2 \sqrt{k^2+m^2} 
\,dk\,, \quad p_{\rm WD}=\frac{1}{3}\frac{2}{(2\pi)^3}  \int_{k=0}^{k_F} 
\frac{4\pi k^2 dk} {\sqrt{k^2+m^2}} \,,
\end{eqnarray}
with $m$ denoting the electron mass and $k_F$ the Fermi wave-number following from the electron number density $n =k_F^3/(3 \pi^2)$. In the non-relativistic limit, $k_F \ll m$, these expressions lead to a polytropic EoS  \cite{Chandrasekhar:1935zz}
\begin{equation}\label{EoSWD}
p= K_{\rm WD} \, \rho^{5/3}\,,
\end{equation} 
with
\begin{equation}
K_{\rm WD } = \frac{(3 \pi^2)^{5/3}}{15 \pi^2 m (\mu_e m_u)^{5/3}} \,,
\end{equation}
$m_u= 1.66 \times 10^{-27}$ kg the unified atomic mass unit \cite{10.1093/ptep/ptaa104}, $\mu_e=A/Z$ the average molecular weight per electron and $A$ and $Z$ the mass and atomic numbers of the core element. For light elements, we have $A=2 \: Z$ and $\mu=2$. Besides, Coulomb effects may be ignored in that case, as they become important when the core of white dwarfs stars consists of heavy elements.

\begin{figure}
    \centering
\includegraphics[scale=.45]{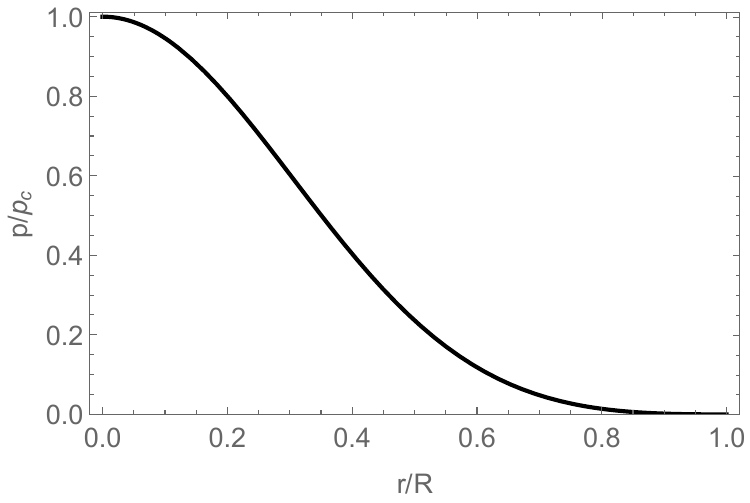}
\includegraphics[scale=.45]{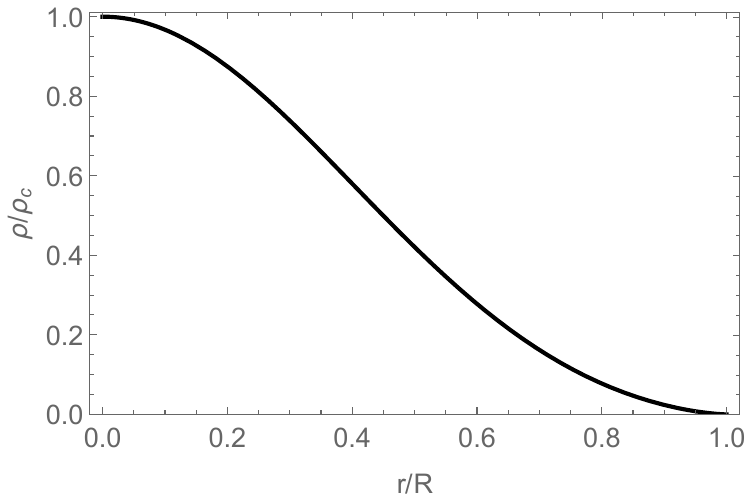}
\includegraphics[scale=.45]{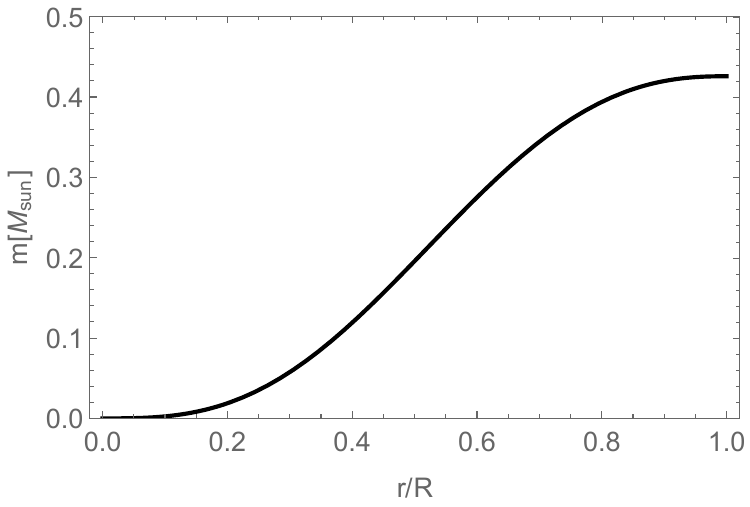}
\includegraphics[scale=.45]{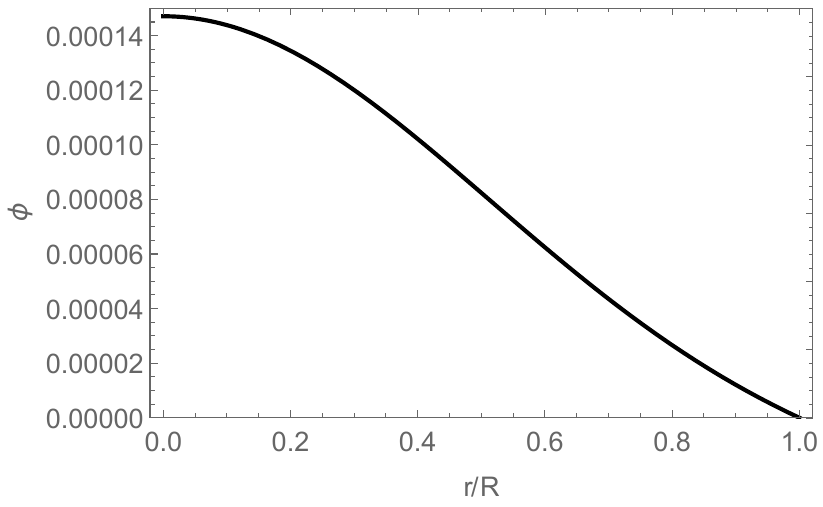}
    \caption{Theoretical energy density, pressure,  mass function and scalar field profile for a white dwarf displaying a polytropic Chandrasekhar EoS \eqref{EoSWD}, a fiducial non-local coefficient $f_2=5$ and a central energy density $\rho_c=7.7 \times 10^5~{\rm gr}/{\rm cm}^3$.} \label{Fig:wd0}
\end{figure}
\begin{figure}
    \centering
    \includegraphics[scale=.6]{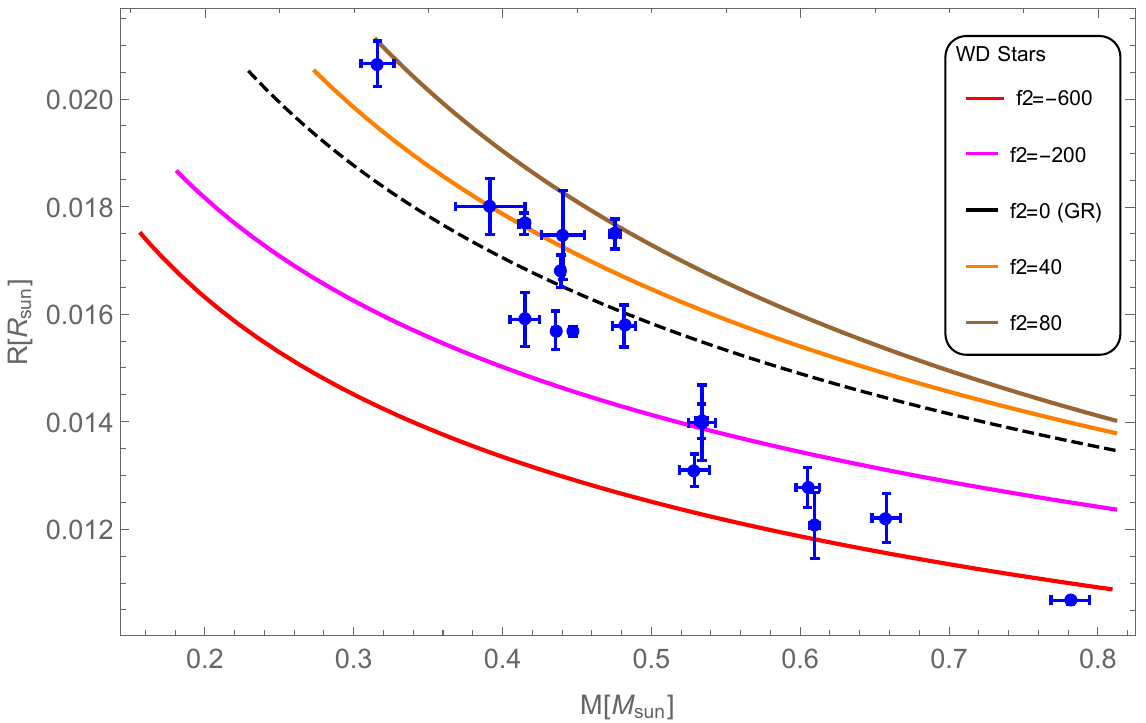}
    \caption{Theoretical white dwarf mass-radius relation for the polytropic Chandrasekhar EoS \eqref{EoSWD} and different values of the non-local coefficient $f_2$. We display profiles for the GR limit $f_2=0$ (black dashed), $f_2=-600$ (red), $f_2=-200$ (magenta), $f_2=40$ (orange), and $f_2=80$ (brown). Positive values of $f_2$ shift the profiles to the right, whereas negative values shift them to the left. The data points are shown for illustration purposes and correspond to observed white dwarfs stars with effective temperatures $T < 20.000$ K \cite{Parsons_2017}, since in that regime the mass-to-radius profile is less sensitive to temperature/core effects \cite{Barstow}}.  \label{Fig:wd}
\end{figure}

In this preliminary analysis, we choose to study the impact of the distorsion function \eqref{fexp} by varying the magnitude of the coefficient $f_2$ and fixing $f_1$ to the value saturating the solar system constraints.~\footnote{We have explicitly verified that varying $f_1$ within the narrow allowed window $\vert f_1\vert \lesssim 5.7 \times 10^{-6}$ \cite{Koivisto:2008dh} does not significantly alter our results.} The typical output of numerically solving the generalized TOV equations \eqref{rr}-\eqref{eq:xiM} for the above EoS is shown in Fig.~\ref{Fig:wd0}, where we display the radial behavior of the pressure, the energy density, the mass function and the field profile for a fiducial value of the coefficient $f_2$ in Eq.~\eqref{fexp}. Note that, as anticipated from dimensional estimates in Section \ref{sec:introduction}, the central value of the $\phi$ field is comparable to the compactness of white dwarfs. The scalar field $\phi$ varies monotonically from $1.4\times 10^{-4}$  at the center to $0$ at the surface. Such a low central $\phi$ value is related to the fact that GR contributions coming from the generalized TOV equations are negligible when computing the structure of white dwarfs. In part, this explains why the internal structure of white dwarfs is still well described within the Newtonian approximation. The mass-radius relation obtained for different $f_2$ values is further depicted in Fig.~\ref{Fig:wd} and compared, for illustration purposes, with the cooler end of white dwarfs stars ($T < 20.000$ K), for which the mass-to-radius profile is less sensitive to temperature/core effects \cite{Barstow}. As clearly illustrated here, large values $f_2\sim {\cal O}(100)$ are \textit{a priori} acceptable within the current uncertainties. The robustness of the precise window $-600 \lesssim f_2\lesssim 80$ following this data comparison should be, however, taken with care, given the potential degeneracies of the $f_2$ parameter with temperature corrections not explicitly accounted for in our analysis. In particular, and as shown explicitly in Ref.~\cite{Parsons_2017}, positive variations of the stellar temperature could shift the theoretical prediction for the radius at a given mass $M$ to higher values with respect to the observational error bars, opening the possibility of obtaining the same $M$-$R$ relation with negative variations of $f_2$. Additionally, one should expect potential bias effects related to our specific choice of the polytropic index in Eq.~\eqref{EoSWD}, which is known to vary from star to star and to be potentially degenerated with the parameters controlling the deviations from the standard Poisson equation \cite{2018JCAP...05..028S}. Since we are mainly interested in a qualitative rather than quantitative description of the impact of nonlocalities in compact stars, we will postpone the analysis of these systematic effects and other related improvements such as the inclusion of leading-order Coulomb corrections \cite{Salpeter:1961zz,1961ApJ...134..683H} or the proper modelization of the envelope thickness and core composition \cite{2018JCAP...05..028S} to future work.

\section{Strange quark stars} \label{sec:strange}

Non-local effects are expected to become important for smaller $f_2$ values in compact objects exceeding the densities of white dwarfs.  Indeed, comparing the quadratic term for a white dwarf in Eq.~\eqref{fexp} with that of an object with compactness $\phi_{\rm C}\gg \phi_{\rm WD}$,  we can easily conclude that they become comparable for 
\begin{equation}\label{estimate}
f^{(C)}_{2} \, \simeq f^{\rm (WD)}_{2} \,\left(\frac{\phi_{\rm WD}}{\phi_{\rm C}}\right)^2 \ll f^{\rm (WD)}_{2} 
\,.
\end{equation}
The recent observation of gravitational waves and electromagnetic radiation from a neutron star merger \cite{GBM:2017lvd, Monitor:2017mdv} illustrates the potential of these objects for constraining modified gravity theories like the Horndeski and beyond Horndeski scenarios  \cite{Ezquiaga:2017ekz,Creminelli:2017sry,Sakstein:2017xjx,Baker:2017hug}. On top of that, it has improved our knowledge of the Quantum Chromodynamics EoS, imposing tight constraints on the maximum mass, radii and tidal deformability of neutron stars \cite{Margalit:2017dij,Bauswein:2017vtn,Rezzolla:2017aly,Ruiz:2017due,Annala:2017llu,Radice:2017lry,Most:2018hfd}. Despite the success, the extraction of relevant information is still rather limited and other less conventional scenarios such as strange-quark stars~\cite{Zhou:2017pha,Drago:2017bnf} remain compatible with data. Given this intrinsic uncertainty, we will restrict ourselves to the latest possibility as a case of study, understanding the corresponding results as qualitatively extensible to neutron star scenarios of similar compactness. 

\begin{figure}
    \centering
    \includegraphics[scale=.46]{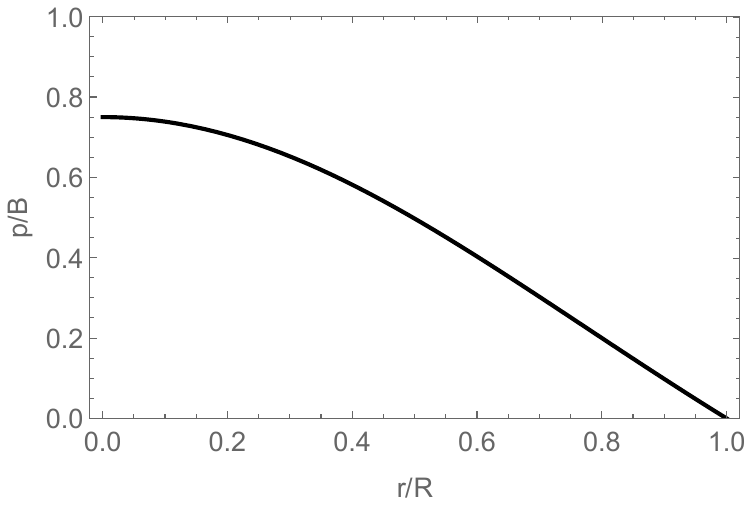}
    \includegraphics[scale=.46]{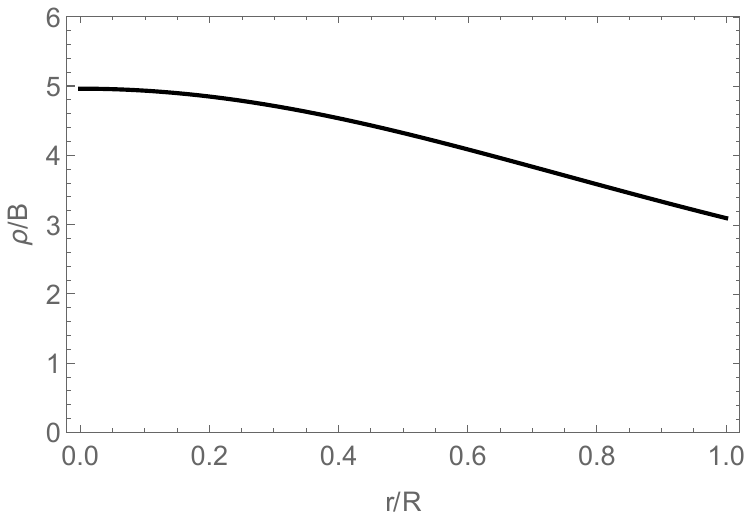}
    \includegraphics[scale=.46]{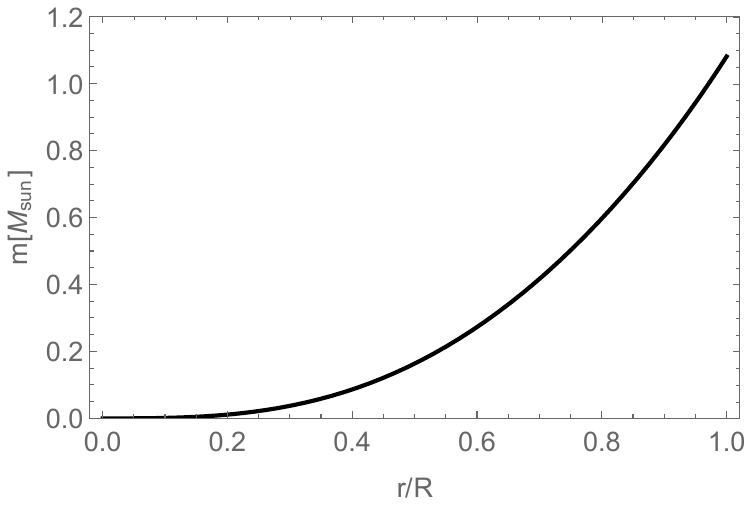}
    \includegraphics[scale=.46]{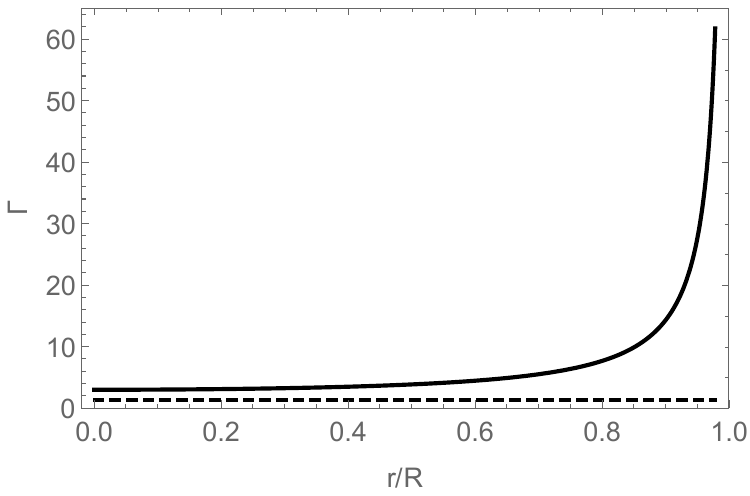}
    \includegraphics[scale=.46]{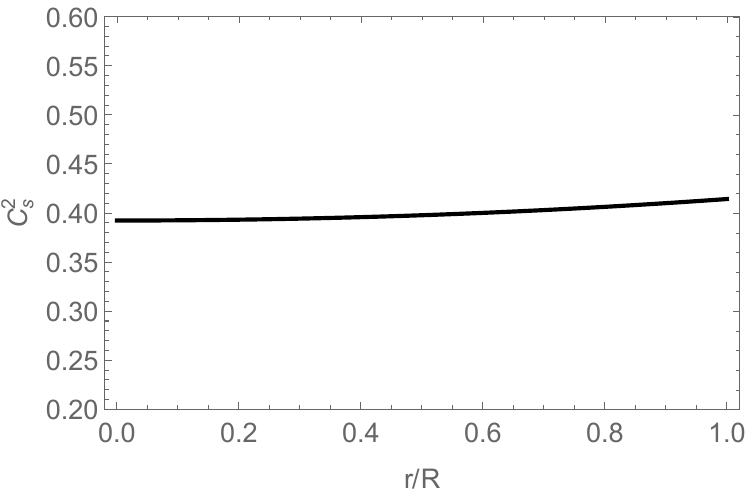}
    \includegraphics[scale=.41]{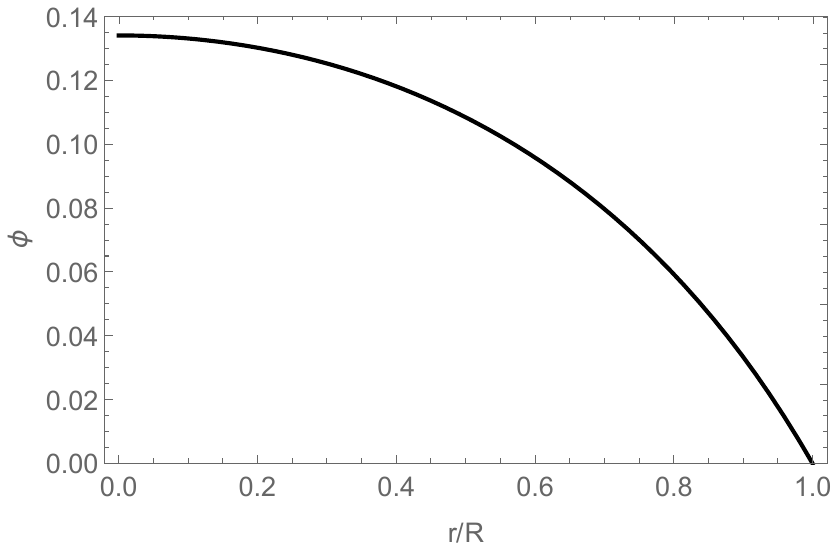}
    \caption{Theoretical energy density, pressure, mass function, scalar field profile, speed of sound and relativistic adiabatic index for strange quark stars displaying the CFL EoS \eqref{EoSstrange}, a fiducial non-local coefficient $f_2=-0.3$ and central pressure $p_c=0.75~B$. Clearly, causality, stability and energy conditions are all fulfilled. 
    }\label{Fig:strange0}
\end{figure}
\begin{figure}
    \centering
    \includegraphics[scale=.7]{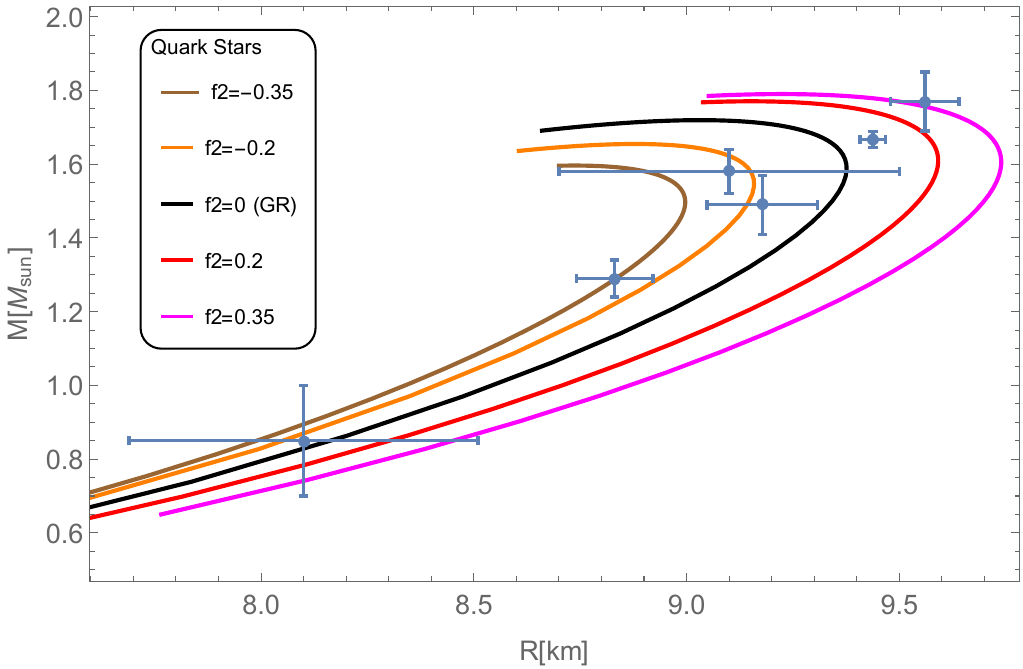}
    \caption{Theoretical strange quark stars' mass-radius relation for the EoS \eqref{EoSstrange} and different values of the non-local coefficient $f_2$. We display profiles for the GR limit $f_2=0$ (black dashed), $f_2=-0.35$ (brown), $f_2=-0.2$ (orange), $f_2=0.2$ (red), and $f_2=0.35$ (magenta). Positive values of $f_2$ shift the profiles to the right, whereas negative values shift them to the left. Data points corresponding to a subset of the strange quark candidates in Ref.~\cite{Aziz:2019rgf} are shown also for illustration purposes.} \label{Fig:strange}
\end{figure}

Strange quark stars are based on the seminal works of Itoh \cite{Itoh:1970uw}, Bodmer \cite{Bodmer:1971we}, Terazawa \cite{Terazawa:1989iw} and Witten \cite{Witten:1984rs}, where it was proposed that strange quark matter consisting of up, down and strange quarks in weak equilibrium could replace $^{56}$Fe as the ground state of Quantum Chromodynamics at asymptotically large densities. According to this idea, the quarks in the stellar interior become effectively massless as compared with the associated chemical potential at very large densities, forming Cooper pairs with a common Fermi momentum. Since those pairs are electrically neutral, electrons cannot be present in this superfluid ground state \cite{Rajagopal:2000ff}, dubbed color-flavor locked (CFL) phase. The associated energy density and pressure at quadratic order in the s-quark mass $m_s$ take the form  \cite{Lugones:2002va} 
\begin{equation}
\rho =  \frac{9 \mu^4}{4 \pi^2 } - \frac{3\, m_s^2\, \mu^2}{4 \pi^2 } + \frac{3}{\pi^2} \Delta^2 \mu^2 + B\,,\hspace{10mm}  p =  \frac{3 \mu^4}{4 \pi^2 } - \frac{3 m_s^2 \mu^2}{4 \pi^2 }
+ \frac{3}{\pi^2} \Delta^2 \mu^2 - B\,, 
\end{equation}
with 
\begin{equation}
\mu^2 = - \alpha + \bigg( \alpha^2 + \frac{4}{9} \pi^2 (\rho - B) \bigg)^{1/2}\,, \hspace{10mm}
\alpha = -\frac{m^{2}_{s}}{6}+\frac{2\Delta^{2}}{3}\,,
\end{equation}
$B$ a phenomenological bag constant encoding the difference between the ''perturbative vacuum" and the true vacuum\footnote{In the MIT bag model \cite{Chodos:1974je,Chodos:1974pn,Farhi:1984qu}, hadrons consist of free or weakly interacting quarks confined to a finite region of space. This region or “bag” is stabilized by adding by hand a term $g_{\mu\nu} B$ to the energy-momentum tensor inside the bag.} and $\Delta$ the superconducting gap. Combining these equations, we get a pressure-density relation \cite{Lugones:2002va}
\begin{equation}\label{EoSstrange}
p= \frac13(\rho- 4B)
+ \frac{2 \Delta^2 \mu^2}{\pi^2} -  \frac{ m_s^2 \mu^2}{2 \pi^2 }\,.
\end{equation}
The first piece in this expression corresponds to the usual MIT bag model (radiation plus constant) for strange quark matter systems \cite{Chodos:1974je,Chodos:1974pn,Farhi:1984qu}. The second one, proportional to $\Delta^2$, is associated with the binding energy of the di-quark condensate and tends to make the system stiffer. The last term, proportional to $m_s^2$, has the opposite effect. 

Since the numerical values of the quantities $m_s$, $B$, and $\Delta$ characterizing the EoS \eqref{EoSstrange} are not accurately known, they will be considered as free parameters in this work. Among the many viable cases  within the stability window, \cite{Flores:2017hpb}
\begin{equation}
m_s^2  <  2 \mu \Delta \,, \hspace{10mm}
B  < \frac{m_n^4}{108 \pi^2} + \frac{m_n^2 \Delta^2}{2 \pi^2} - \frac{m_s^2 m_n^2}{12 \pi^2}\,,
\end{equation}
with $m_n \simeq 939~{\rm MeV}$ the neutron mass, we shall consider here a scenario with $B=120\, \textrm{MeV}\cdot \textrm{fm}^{-3},$ $m_s=150\, \textrm{MeV}$ and $\Delta=150\, \textrm{MeV}$, in agreement with other phenomenological studies suggesting that $B > 57~{\rm MeV}/{\rm fm}^3$ \cite{Farhi:1984qu} and $\Delta=(100-200)~{\rm MeV}$ \cite{Baym:2017whm}. This corresponds to the CFL18 model in Table I of Ref.~\cite{Flores:2017hpb}.

The output of numerically solving the generalized TOV equations \eqref{rr}-\eqref{eq:xiM} for this set of parameters and a fiducial value $f_2=5$ is displayed in Fig.~\ref{Fig:strange0}. Like in the case of white dwarf stars, the scalar field $\phi$ is a decreasing function of the radial coordinate from the center to the surface of the strange quark star, varying from $1.4\times 10^{-1}$ to $0$. However, in the latter class of stars, the central value of $\phi$ is three orders of magnitude larger than in the case of a white dwarf, being this enhancement related to the larger compactness of strange quark stars and the associated increase of nonlocal contributions in the TOV equations. The pressure and the energy density of the strange star exhibit the typical behavior of quark matter. In particular, while both of them decrease monotonically with the radial coordinate, only the pressure vanishes at the surface of the star. Note also that the physicality conditions \eqref{cs}, \eqref{gamma} and \eqref{SEC} are clearly satisfied throughout the system.

Varying the magnitude $f_2$ for a fixed $f_1$ saturating the solar system bounds, we obtain the mass-radius relation depicted in  Fig.~\ref{Fig:strange}, where we display also the ensemble of the data points of Ref.~\cite{Aziz:2019rgf} for illustration purposes. As expected from the largest compactness of these objects, cf.~Eq.~\eqref{estimate}, the allowed values for $f_2$ turn out to be significantly smaller than those obtained from white dwarfs, being roughly described by a restricted set of ${\cal O}(1)$ values. As before, the precise window $-0.35 \lesssim f_2\lesssim 0.35$ following this data comparison should be understood just an order of magnitude estimate, since its two extremes depend implicitly on the precise equation-of-state (CFL18) and data subset involved in the modeling. In particular, a positive/negative variation of bag constant $B$ in Eq.~\eqref{EoSstrange} could be potentially compensated by a positive/negative variation of $f_2$, as seen from the sensitivity of the $M$-$R$ relation in Ref.~\cite{Aziz:2019rgf}. For instance, for the CFL12 model in Table I of Ref.~\cite{Flores:2017hpb}, $B=100\, \textrm{MeV}\cdot \textrm{fm}^{-3},$ $m_s=0$ and $\Delta=100\, \textrm{MeV}$ and the aforementioned range on $f_2$ for the same data subset get shifted to $-0.8 < f_2 < 0.2$.  

\section{Conclusions}\label{sec:conclusions}

We have argued that compact stars comprise an interesting laboratory to test the imprints of non-local theories of gravity at small scales, being potentially complementary to solar system tests. To illustrate this, we have considered a simple parametrization of non-localities involving an analytic distortion function $f(\Box^{-1} R)$. After recasting the associated Lagrangian density in a local representation, we made use of a static and spherically symmetric ansatz to reduce the equations of motion to a generalized Tolman-Oppenheimer-Volkoff form. The impact of non-localities on compact astrophysical objects has been estimated by considering two physically-motivated equations of state: i) a Chandrasekhar model describing white dwarfs, and ii) a color-flavor locked phase describing quark matter in very dense strange stars, understanding the latest as a proxy of neutron stars in terms of compactness. In both cases, we have obtained well-behaved metric solutions satisfying causality and stability criteria and able to describe realistic astrophysical configurations. Our results confirm the naive expectation: the typical value of the scalar field at the center of strange quark stars is significantly larger than the one obtained in the case of white dwarfs, in agreement with the larger compactness of those objects. This implies that, in spite of the many uncertainties involved, the analysis based on strange quark stars gives rise to qualitatively tighter constraints on non-localities, at least for the data sets considered here. Specifically, like the magnitude of the central scalar field value in both stars, the constraints in a quark star are three orders of magnitude better than those coming from a white dwarf. It would be interesting to extend this study to neutron stars, even though we expect to obtain a similar order of magnitude range for $f_2$. In this case, the star consists of a core with a hadronic equation of state and a crust with a polytropic equation of state, meaning that proper modelling of the system should take into account, among other aspects, the phase transition at the boundary between these two layers. The existing data sets on the maximum mass configurations on neutron stars provide an appealing way of restricting the parameter $f_2$. We hope to be able to address this challenging computation in a future publication.

The quantitative results presented in this paper should be of course taken with care, given the current theoretical and observational uncertainties on the defining properties of these compact stars. Note, however, that this line of research is expected to grow in the near future, given the significant amount of data that will be available from the next generation of astronomical surveys. On the one hand, recent models of stellar populations suggest that galaxies like the Milky Way contain about ten billion white dwarfs \cite{2009JPhCS.172a2004N}. The GAIA mission alone has very likely discovered already more than 260.000 white dwarfs in the Solar neighbourhood \cite{2019MNRAS.482.4570G}, making these compact stars potential laboratories for testing alternative theories of gravity. On the other hand, the upcoming X-ray and radio surveys \cite{2019BAAS...51c.425F} will also increase considerably the number of neutron and potential strange quark stars observed, particularly in the central region of the Milky Way \cite{2018ASPC..517..793B}, where the number of compact objects is expected to be significant. 

\section*{Acknowledgments}
 
We thank the Funda\c c\~ao para a Ci\^encia e Tecnologia (FCT), Portugal, for the financial support to the Center for Astrophysics and Gravitation-CENTRA, Instituto Superior T\'ecnico,  Universidade de Lisboa, through the Project No.~UIDB/00099/2020. G.~P. and I.~L. acknowledge also the support of this agency through the grant No. PTDC/FIS-AST/28920/2017. JR is supported by a Ram\'on y Cajal contract of the Spanish Ministry of Science and Innovation with Ref.~RYC2020-028870-I. He thanks also Anupam Mazumdar for useful discussions on the landscape on non-local theories of gravity and the Fundação para a Ciência e a Tecnologia (Portugal) for financial support through a CEECIND/01091/2018 grant during the first stages of this work. 

\bibliographystyle{JHEP.bst}
\bibliography{biblio.bib} 

\end{document}